\documentstyle[12pt]{article}

\begin{document}
\setcounter{page}{1} \pagestyle{plain} \vspace{1cm}
\begin{center}
\Large{\bf Generalization of the Randall-Sundrum Model Using Gravitational Model $F(T,\Theta)$}\\
\small \vspace{1cm} {\bf S. Davood Sadatian$^1$}~and~ {\bf S. M.
Hosseini}\\
\vspace{0.5cm} {\it Department of Physics, Faculty of Basic
Sciences,University of Neyshabur, P. O. Box 9318713331, Neyshabur,
IRAN}\\
$^1${\it sd-sadatian@um.ac.ir}
\end{center}
\vspace{1.5cm}
\begin{abstract}
In this letter, we explore a generalized model based on two
scenarios including the Randall-Sundrum model and Gravity model
$F(T,\Theta)$. We first study the standard Randall-Sundrum
Gravitational model and then add a function containing two
parameters as torsion and trace energy-momentum tensor to the main
action of the model. Next, we derive the equations of the
generalized model and obtain a new critical value for the energy
density of the brane. The results showed that inflation and the dark
energy dominated stage can be realized in this model.
\end{abstract}
\newpage

\section{Introduction}
According to the standard model of cosmology, the Planck Era is
referred to a time starting from the creation of the Universe (about
14 billion years ago) until $10^{-43}$ seconds. In this time-lapse,
quantum effects are important parameters that can be formulated
using the quantum gravity theory [1]. One of the ideas for
describing quantum gravity is the string theory [2]. Some features
of this theory include the replacement of one-dimensional quantum
operators with pseudo-point operators in Hilbert space and the need
for the presence of supersymmetry instead of symmetry [3]. The idea
of the cosmology of the string emerged in the early 1990s by
extending the string idea into the cosmological model [4]. According
to this theory, from the mathematical point of view, we need a space
with higher dimensions. The outcome of this idea is that the
fundamental constants of the physics have a variable form with
respect to time. The image of these constants in a (3 + 1)
dimensional world of our space-time universe is constant with
respect to time [5].\\
Kaluza (1921) and Klein (1926) proposed the unified field theory of gravitation and electromagnetism,
which needs an extra spatial dimension that is a function of the Planck length [6].\\
Randall and Sundrum [7] presented a model in which the Universe is a
brane embedded in the bulk. In other words, the universe is located
in a less spatial dimension than the main dimension of the bulk with
five-dimensional space-time [8]. The main feature of the bulk is the
gravity which can freely propagate in it. Because only a graviton
particle can exist inside the bulk in this model, the other
particles of the standard model are in the braneworld. Based on the
number of branes available in this model, two Randall-Sundrum (RS)
models (I) and (II) have been presented [7,8]. The five-dimensional
space-time determines the type of model. In type (I) Randall model,
the distance  $L$ is limited while in the type (II) Randall model
$L$ tends to infinity.
The adjustable parameters of this model include cosmic bulk constant and energy density (tension) of the brane [7-9].\\
In this model, the torsional metric in the five-dimensional
space-time is defined according to Eq. (1):
\begin{equation}
ds^2=e^{-2A|y|}\eta_{\mu\nu}dx^\mu dx^\nu+dy^2
\end{equation}
where $ 0\leq y\leq \pi r_c$ is the fifth dimension of space and
$r_c$ is essentially a compactification "radius" [7].\\
Given the above explanations, we first study the Randall-Sundrum
standard model. Then, we obtain the equations of the model by
generalizing this model by adding a function $F(T,\Theta)$ to the
main action of the model. Finally, we determine some cosmological
parameters by our generalized model equations.\\

\section{Teleparallel gravity model}
The equations of the gravitational field in general relativity are
based on the interaction between the curvature of space-time and the
energy density content of the universe. Other models have been
proposed for gravity [1-5] by generalizing a gravity model based on
scalar and Ricci tensor (curvature). One of these models is based on
the replacement of the torsion
tensor rather than the curvature of space-time in the gravity model [10]. This modified model proves the inflation at the beginning of the universe.\\
In the following, a brief overview of the above model is presented.
We assume Planck constant, Boltzmann constant, and the speed of
light as $\hbar=K_B=c=1$ and define the gravitational constant $8\pi
G$ by the relation $K^2 =\frac{8\pi}{M^2_{pl}}$ (where $M_{pl}$ is
Planck mass). In this model, the torsion tensor is denoted by the
relation $T^\rho_{~\mu\nu}=r^\rho_{~\nu\mu}-r^\rho_{~~ \mu\nu}$,
where $r^\rho_{~\nu\mu}=e^\rho_A\partial_\mu e^A_\nu$ is the
connection without the Weitzenbock curvature[8]. The contortion
tensor is also defined by the relation $K^{\mu\nu}_{
~~\rho}=\frac{1}{2}(T^{\mu\nu}_{~~\rho}-T^{\nu\mu}_{~~\rho}-T^{
~~\mu\nu}_\rho)$. In addition, the torsion scalar is presented as
$T=S^{~\mu\nu}_\rho K^{~\rho}_{\mu\nu}$ using the super potential
$S^{~\mu\nu}_\rho$ and the contortion tensor $K^\rho_{~\mu\nu}$. The
action of this modified model with the matter in term of $F(T)$  is
shown by Eq. (2) [10]
\begin{equation}
S=\int d^4x|e|(\frac{F(T)}{2K^2}+L_M)
\end{equation}
where $|e|=det(e^A_\mu) =\sqrt{-g}$ and $L_M$ is the Lagrangian of
matter. Regardless of material Lagrangian, gravity action in five
dimensions is defined by the following relations
\begin{equation}
(^{(5)}S)=\int d^5x|^{(5)}e|(\frac{F(^{(5)}T)}{2K^2_5})
\end{equation}

\begin{equation}
(^{(5)}T)=\frac{1}{4}T^{abc}T_{abc}+\frac{1}{2}T^{abc}T_{cba}-T^{~~a}_{ab}T^{cb}_{~~c}
\end{equation}
where $^{(5)}e=\sqrt{^{(5)}g}$ and the gravitational constant and
Planck mass in five dimensions are related to each other by the
relation $K^2_{5}=8\pi G_5=\frac{8\pi}{M^2_{pl}}$ [10].

\subsection{Compression of Kaluza-Klein}
In this subsection, we need to explain the compression mechanism of
the Kaluza-Klein [6]. In these five dimensions, the metric is
represented by the following diagonal matrix:
\begin{equation}
^{(5)}g_{ab}=\left[ \begin {array}{cc} {\it g_{\mu \nu}} &0\\
\noalign{\medskip}0&{- \varphi }^{2}\end {array} \right]
\end{equation}
where the scalar field is uniform and dependent on time according to
the relation $\varphi^2=\Re^2\theta^2$, Here, $\varphi$ is related
to $\Re$(compressed space radius) and the quadratic orthogonal
components are represented in a compact space dimension with a
dimensionless characteristic $\theta$ [10]. According to the
relationship $g_{\mu\nu}=\eta_{AB}e^A_{~\nu}e^B_{~\nu}$, the second
order metric tensors in five dimensions and tangent spaces in five
dimensions are defined according to the relations $ \eta_{AB}=
diag(1,-1,-1,-1,-1)$ and $e^A_{~\nu}=diag (1,1,1,1,\varphi)$,
respectively. Therefore, using these relations in Eqs. (3) and (4),
the effective action is given by the compression mechanism of the
Kaluza-Klein as follows:
\begin{equation}
S=\int d^4x\frac{1}{2K^2}\varphi |^{(4)}e|
F(T+\varphi^{-2}\partial_\mu\varphi\partial^\mu\varphi)
\end{equation}
where the shape of the gravitation function and  $
|^{(5)}e|=|e|\varphi$ does not change in comparison with the
relation
(3). This function contains a torsion scalar component and a scalar field.\\
In the simplest case, assume a gravitational function in form
$F(T)=T-2\Lambda_4$, where $\Lambda_4$ is a positive cosmological
constant. If we define a scalar field $\sigma$ in the form of
$\varphi=\sigma^2\xi$, where $\xi=\frac{1}{4}$,by rewriting Eq. (6),
the effective action is obtained as follows:
\begin{equation}
S^{eff}_{KK}=\int
d^4x\frac{1}{K^2}|^{(4)}e|[\frac{1}{8}\sigma^2T+\frac{1}{2}\partial_\mu
\sigma\partial^\mu \sigma-\Lambda_4]
\end{equation}
where the metric in the flat universe of FLRW is defined as
$ds^2=dt^2-a^2(t)\Sigma _{i=1,2,3} (dx^i)^2$, $a$ is the scale
factor, and $H=\frac{\dot{a}}{a}$ is the Hubble parameter. Moreover,
the vector of the space tangent and metric are defined with the
relations $e^A_{\mu}=diag(1,a,a,a)$ and
$g_{\mu\nu}=diag(1,-a^2,-a^2,-a^2)$, respectively. If $T=-6H^2$, the
equations of the gravity field are calculated by the following
equations[10].
\begin{equation}
\frac{1}{2}\dot{\sigma}^2-\frac{3}{4}H^2\sigma^2+\Lambda_4=0
\end{equation}
\begin{equation}
\dot{\sigma}+H\sigma\dot{\sigma}+\frac{1}{2}\dot{H}\sigma^2=0
\end{equation}
And the equation of motion for the scalar field is obtained by Eq.
(10):
\begin{equation}
\ddot{\sigma}+3H\dot{\sigma}+\frac{3}{2}H^2\sigma=0
\end{equation}
By combining the above gravity field equations, Eq. (11) is
obtained:
\begin{equation}
\frac{3}{2}H^2\sigma^2-2\Lambda_4+H\sigma\dot{\sigma}+\frac{1}{2}\dot{H}\sigma^2=0
\end{equation}
With solving this equation in terms of H, the obtained  value ($H$)
is achieved corresponding to the Hubble parameter in the inflation
time [10].
\subsection{Effective gravity in Randall-Sundrum model}
Considering the solution of vacuum equations in a five-dimensional
space-time and (RS) type (II) model, which has only one mass within
a five-dimensional bulk with positive energy density, it is
concluded that the five-dimensional space-time is an Anti deSitter
(AdS) space. In the following, according to the Randall model type
II [11-12] in the Teleparallel Gravity theory, Friedmann equations
for the brane in the FLRW background metric for effective gravity
$F(T)$ are determined as follows
\begin{equation}
H^2\frac{dF(T)}{dT}=-\frac{1}{12}[F(T)-4\Lambda-2K^2\rho_M-(\frac{K^2_5}{2})Q\rho^2_m]
\end{equation}
where $Q=(11-60\omega_M+93\omega^2_M)/4$. In this relation
$\omega_M=\frac{P_M}{\rho_M}$ corresponds to the perfect fluid
equation of state for the pressure and density of the confined
matter in the brane. $\Lambda=\Lambda_5+(K^2_5/2)^2\lambda^2$ with
$0<\lambda$, brane tension, and $G=[1/(3\pi)](K^2_5)^2\lambda$. With
substituting $F(T) =T-2\Lambda_5$ in Eq. (11), the approximate
solution of de sitter for brane is obtained as
$H=H_{DE}=\sqrt{(\Lambda_5+K^4_5\lambda^2/6)}$, where $a(t)=
a_{DE}e^{(H_{DE}t)}$ with $a_{DE}>0$, and $T=-6H^2$ is assumed.
Therefore, this relation can describe the accelerating expansion of
the universe [10].

\section{Teleparallel gravity model with a trace of momentum-energy
tensor} Another example of the generalized model based on the $F(T)$
model that we discuss here is the $F(T,\Theta)$  model.\\
An important feature of this model is the effect of geometry and the
universe content through space-time torsion and trace of
momentum-energy tensors, where $T$ is torsion scalar and $\Theta$ is
energy-momentum tensor trace. Therefore, in order to study this
model, we consider the proposed gravity function  $F(T,\Theta)$ in a
five-dimensional space-time compressed by the Kaluza-Klein theory
[10,13].\\
\begin{equation}
F(T,\Theta)=\alpha(-T)^n(\Theta)^m\tanh(\frac{T_0}{T})
\end{equation}
where  $\alpha=1, m=1, n=2, T_0=-6H^2_0$. \\
First, we transfer the five-dimensional space-time in which the
gravity model is related to the scalar field, and coupled with the
additional dimension of space, into a four-dimensional space-time in
a brane with a flat metric FLRW using the compression mechanism of
the Kaluza-Klein. We have already investigated that in the gravity
model $F(T)$ in the four-dimensional space we get the Friedmann
equations according to Eq. (12)[10]. Therefore, due to the
characteristics of the gravity model $F(T,\Theta)$  mentioned above,
we rewrite Friedmann equations for this gravitational model. In this
model, the torsion is based on the relation $T=-6H^2$  and $\Theta$
is the momentum-energy tensor trace. The total momentum-energy
tensor on brane is written according to Eq. (14)[12]:
\begin{equation}
T^{A}_B=S^{A}_B\delta(y)
\end{equation}
where $S_B^{A}=diag (-\rho_b,P_b,P_b,P_b,0)$ and $P_b,\rho_b$ are
pressure and density of the total brane energy, respectively. By the
conjugation condition in $y=0$ in the five-dimensional space and
simplify the relations, we have:
\begin{equation}
H^2=(\varepsilon
\frac{\rho^2_b}{36}\chi^4)+\frac{\Lambda}{6}-\frac{k}{a^2}+\frac{C}{a^4}
\end{equation}
where $C$ is constant of integration. It is of note that this
relation is established in the brane and the energy of conservation
law will be conserved according to the following relation.
\begin{equation}
\dot{\rho_b}+3H(\rho_b+P_b)=0
\end{equation}
Assuming that $\rho_b=\rho+\lambda$, where $\lambda$ is a brane
tension and
$\Lambda=\varepsilon\lambda^2\frac{\chi^4}{6}$($\varepsilon=1$, if
the extra dimension is space-like, while $\varepsilon=-1$, if it is
time-like), we get the following relation in a flat metric $(k=0)$,
where $C=0$ is the constant of integration and called dark
radiation:
\begin{equation}
H^2=(\frac{\rho}{3})(1+\frac{\rho}{2\lambda})
\end{equation}
Now, by applying the gravity function $F(T,\Theta$), Eq. (13), in
the Friedman equation obtained from the compression of space KK, Eq.
(12), we have Eq. (18):
\begin{equation}
\tanh(\frac{T_0}{T})[A(-T)^n+BT^{n-1}+C(-T)^{n+1}]=\frac{8\pi
G}{12(3\omega-1)}+\frac{\pi
G}{16\lambda(3\omega-1)}(11-60\omega+93\omega^2)\rho
\end{equation}
where $A=-\frac{n}{6} , B=\frac{T_0}{6} , C=\frac{1}{72}$ . Assuming
$n=2$ and due to the relation of momentum-energy tensor (14) we get
$\Theta=3P-\rho$, which can be rewritten as $\Theta= (-1+3\omega)$
using the equation of state. Using the Maclaurin expansion for the
hyperbolic tangent and use the first-order approximation we have:
\begin{equation}
{T}=\frac{T_0}{72}[\frac{8\pi G}{12(3\omega-1)}+\frac{\pi
G}{16\lambda(3\omega-1)}(11-60\omega+93\omega^2)\rho]-12
\end{equation}
Here, assuming that $\rho_b=\lambda\sqrt{\rho}$ and substituting it
in Eq. (14), we get Eq. (18). Moreover, the Hubble parameter is
obtained according to Eq. (20) using the relation $T=-6H^2$.
Therefore, we have:
\begin{equation}
{H}=\pm(-\frac{T_0}{432}[\frac{8\pi G}{12(3\omega-1)}+\frac{\pi
G}{16\lambda(3\omega-1)}(11-60\omega+93\omega^2)\rho]+2)^{\frac{1}{2}}
\end{equation}
Using Eq. (20) and assuming that $\rho_b=\lambda\sqrt{\rho}$, the
critical value for the energy density of the brane is determined
according to the following equation:
\begin{equation}
\rho_b<\sqrt{\frac{863\Theta \lambda^3}{\pi GT_0}}
\end{equation}
Now, we can obtain the universe scale factor from the combination of
Hubble parameter and relation (20), for the inflation period using
the hyperbolic tangent series and the first-order approximation.
Also, using an approximate of the deSitter solution on the brane (
assuming $T_0=-6H_0^2=-24\times10^{-36}$ ), we obtain scale factor
$a(t)$, shown in Fig. 1.
\begin{figure}[htp]
\begin{center}\includegraphics{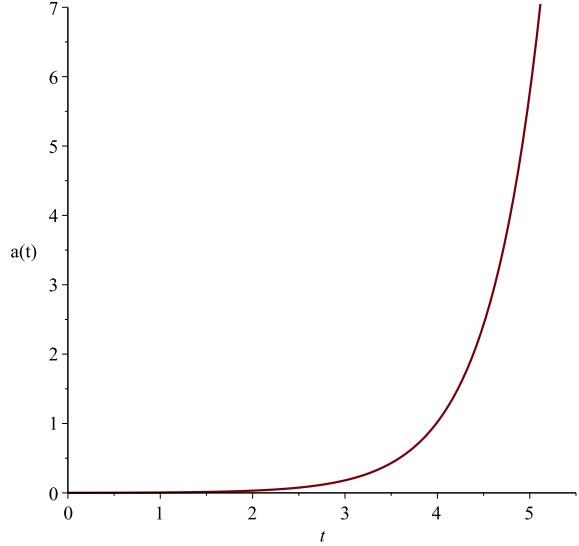} \vspace{6cm}\
\end{center}
\caption{\small { The universe scale factor for the inflation
period(where $\alpha=1, m=1, n=2$.)}}
\end{figure}
To calculate the equation of state $\omega(t)$, it is necessary to
solve simultaneously three nonlinear differential equations
according to the obtained relations for the gravity model
$F(T,\Theta)$ in the five-dimensional space. For this purpose,
first, by combining Eq. (16) and the equation of state $p=\omega\rho
c^2$, we get the differential equation between the energy density of
the brane and  $\omega(t)$ as follows:
\begin{equation}
\dot{\rho_b}+3H\rho_b(1+\omega(t))=0
\end{equation}
Now, according to the pressure and energy relation for the
Teleparallel gravity model [14] and combining it with the equation
of state, we get the nonlinear differential equation for the scalar
field and the Hubble parameter as follows:
\begin{equation}
(\frac{\dot{\varphi}}{2}-\frac{3}{4}\varphi^2H^2+\Lambda_4)\omega=\frac{\dot{\varphi}^2}{2}+H\varphi\dot{\varphi}+\frac{3}{4}H^2\varphi^2+\frac{1}{2}\varphi^2\dot{H}-\Lambda_4
\end{equation}
By solving Eqs. (22) and (23) simultaneously and also the equation of motion for the scalar field (10),
the equation of state $\omega(t)$ is obtained.
In the following, we used the first-order approximation to solve differential equations.
Moreover, by adjusting the brane tension parameter, the energy density and the potential of the scalar field with the cosmological constant on the brane,
we get the equation of state in terms of time that shown in Fig. 2. As shown in this model, the phantom boundary crossing occurred [15,16].\\
\begin{figure}[htp]
\begin{center}\includegraphics{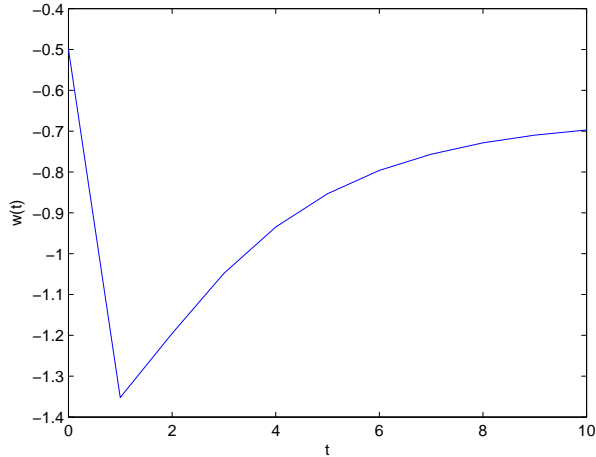} \vspace{5cm}\
\end{center}
\caption{\small {The equation of state $\omega(t)$ in terms of time
for generalized RS model with $F(T,\Theta)$ for $\alpha=1, m=1,
n=2$}}
\end{figure}
It is shown that inflation or the dark energy dominated stage can be
realized only by the effect of the torsion and trace energy-momentum
tensor without the curvature. As a result, it can be interpreted
that these models may be equivalent to the Kaluza-Klein and RS
models without gravitational effects of the curvature but just due
to those of the torsion and trace energy-momentum tensor in
teleparallelism. Indeed, this is the new work on the concrete
cosmological solutions to describe the cosmic accelerated expansion
of the KK and RS models in $F(T,\Theta)$  gravity. Based on this
results, it can be stated that phenomenological $F(T,\Theta)$
gravity models in the four-dimensional space-time can be derived
from more fundamental theories. In this regard, the observational
constraints on the derivative of $F(T)$  and similar function as
$F(T,\Theta)$ until the fifth order were presented in Ref. [17] with
cosmographic parameters acquired from the observational data of
Supernovae Ia and the baryon acoustic oscillations. The results of
the model presented in this work, as a concrete example of
$F(T,\Theta)$ gravity models, are consistent with those obtained in
[17].\\
\section{Conclusion}
In this paper, a generalized gravity model was proposed based on the
time-space torsion and interaction with the universe content. The
study of this gravity model in a five-dimensional space according to
the Randall-Sundrum approach included a four-dimensional brane in a
five-dimensional bulk. In this regard, the Kaluza-Klein theory was
used to compress the fifth dimension of space in this gravity model.
Then, in accordance with the inflation period of the standard
cosmological model, the new critical value for the energy density of
the brane, the Hubble parameter, and the scale factor were obtained.

\end{document}